\documentclass{ws-procs961x669}            
\usepackage{cancel,mathtools,physics,tensor,mathrsfs,dsfont,comment,enumerate,enumitem,cases,wasysym,simplewick}

\def\cA{{\mathcal A}}

\def\cH{{\mathcal H}}

\def\cM{{\mathcal M}}

\def\cS{{\mathcal S}}

\def\cW{{\mathcal W}}

\def\beq{\begin{eqnarray}}
\def\eeq{\end{eqnarray}}
\def\pa{\partial}
\def\at{\left(}               		   
\def\ag{\left\{}              		   

\def\ct{\right)}              		   
\def\cg{\right\}}             		   
\def\d{\mathrm{d}}
\def\e{\mathrm{e}}

\newcommand{\wick}[1]{{:}#1{:}}
\newcommand{\Eq}{Eq.\@ }
\newcommand{\Eqs}{Eqs.\@ }

\newcommand{\expvalom}[1]{\expval{\wick{#1}}_\omega}
\newcommand\eqH{\stackrel{\mbox{\tiny $\cH$}}{=}}
\newcommand\leqH{\stackrel{\mbox{\tiny $\cH$}}{<}}

\begin{document}
\title{Trace anomaly and evaporation of spherical black holes}

\author{P.Meda$^*$}

\address{Dipartimento di Fisica, Universit\`a degli Studi di Genova,  \\
Istituto Nazionale di Fisica Nucleare - Sezione di Genova,\\
Genova, 16146, Italia\\
$^*$E-mail: paolo.meda@ge.infn.it}

\begin{abstract}
The evaporation of four-dimensional spherically symmetric black holes is presented in the framework of quantum field theory in curved spacetimes and semiclassical gravity. It is discussed how the evaporation process can be sourced by the presence of the trace anomaly of a massless, conformally coupled scalar field outside the apparent horizon of the black hole.
\end{abstract}


\bodymatter

\section{Quantum fields in curved spacetimes}

From the time of the discovery made by Hawking that black holes can emit radiation \cite{Hawking74bh,Hawking75creat}, black hole evaporation has become of great interest in analyzing the interplay between matter and gravity. A theoretical framework where modelling black hole evaporation is represented by quantum field theory in curved spacetimes \cite{Hollands15qfcs}, which consists of describing the propagation of a quantum field over a classical curved background. On one side, the background is assumed to be a four-dimensional globally hyperbolic spacetime $(\mathcal{M},g)$, where $\mathcal{M}$ is a smooth manifold and $g$ a Lorentzian metric with signature $(-,+,+,+)$. On the other side, the quantum field is the fundamental observable, which is viewed as an operator-valued distribution living in a $*$-algebra $\cA(\cM)$. For a real, free Klein-Gordon field $\phi$ the $*$-algebra is generated by the set of linear smeared fields $\{\phi(f),f\in\mathcal{D}(\mathcal{M})\}$ satisfying the following relations:
\[ 
	\phi({P} f)=0, \qquad \phi(f)^{*}=\phi(\bar{f}), \qquad[\phi(f), \phi(h)]=\mathrm{i} \Delta(f, h).
\]
Here, $P = -\square_g + m^2 + \xi R$ is the Klein-Gordon operator constructed from the d'Alambertian operator $\square_g = g^{\mu\nu} \nabla_\mu\nabla_\nu$ on $(\mathcal{M},g)$, and $\Delta=\Delta_R-\Delta_A$ is the causal propagator of the spacetime, which is defined as the difference between the unique retarded and advanced fundamental solutions of $P$. A generic quantum state $\omega$ is a linear, positive, normalized functional over $\cA(\cM)$, and it is determined by the $n$-point functions
\[
	\omega_n(f_1,\dots, f_n)  = \expval{\phi(f_1)\dots \phi(f_n)}_\omega \in\mathcal{D}'(\mathcal{M}^n).
\]
Actually, for Gaussian pure states it is sufficient to known the kernel of the two-point function $\omega_2(x',x) = \expval{\phi(x')\phi(x)}_\omega$. 

Different from the flat case, where the Poincar\'{e} invariance selects a vacuum state uniquely as preferred reference state, in a curved spacetime there is not in principle a preferred notion of ``vacuum'' dictated by the symmetries of the spacetime. However, a guiding principle to select physically reasonable states in curved spacetimes is that Wick normal-ordered fields like $\wick{\phi^2}$ and $\wick{T_{\mu\nu}}$ should acquire finite expectation values when evaluated on $\omega$. In globally hyperbolic spacetimes, the set of Hadamard states represents the right class where computing normal-ordered fields, because they share the same ultraviolet singular structure of the Minkowski vacuum two-point function $\expval{\phi(x')\phi(x)}{0}$. Given $\sigma(x',x)$ the one half of the signed geodesic distance between $x'$ and $x$, the two-point function of a Hadamard state is always locally of the form
\[ 
	\omega_2(x',x) = \frac{1}{8\pi^2} \at \mathscr{H}_{0^+}(x',x) + w(x',x) \ct.
\] 
Here,
\begin{equation}
\label{def:Hadamard}
	\mathscr{H}_{0^+}(x',x) = \lim_{\varepsilon \to 0^+}\frac{u(x',x)}{\sigma_{\varepsilon}} + \sum_{n \geq 0} v_n(x',x) \sigma(x',x)^n \log \left(\frac{\sigma_{\varepsilon}}{\mu^{2}}\right)
\end{equation}
is the universal singularity called Hadamard parametrix viewed in the distributional sense, where $\sigma_{\varepsilon}(x',x)= \sigma(x',x) + \mathrm{i}\epsilon( t(x')-t(x))$, $t$ is any time function, and $\mu > 0$ is a length scale. The coefficients $u(x',x)$, $v_n(x',x)$ are real-valued bi-scalars fixed by the geometry and the equation of motion, while $w(x',x)$ characterizes the state, and it must be chosen in such a way that $\omega_{2}$ is a bi-solution of the Klein-Gordon equation. The Hadamard point-splitting regularization consists of subtracting the divergences contained in $\mathscr{H}_{0^+}(x',x)$ before computing the coinciding point limits: it generalizes the normal-ordering prescription in the flat spacetime, and it is also local and covariant, because only local geometry enters in the subtraction. This procedure, and more generally any prescription to obtain Wick observables, is always unique up to a fixed combination of local terms depending on the mass and on the curvature, whose coefficients represent the renormalization freedom of the quantum theory \cite{Hollands01wick,Hollands05stress}.

\subsection{Semiclassical gravity, the quantum stress-energy tensor, and the trace anomaly}

Strictly related to a formulation of quantum fields propagating in curved backgrounds is the semiclassical theory of gravity, which aims to analyze the back-reaction of quantum fields on the spacetime geometry using the semiclassical Einstein equations
\begin{equation}
\label{eq:SCE}
	G_{\mu\nu} = 8\pi \expvalom{T_{\mu\nu}}
\end{equation}
in units convention $G = c = \hbar = 1$. In this equations, the classical Einstein tensor is equated to the renormalized quantum stress-energy tensor evaluated in a certain quantum state $\omega$. Any pair formed by a spacetime and a quantum state satisfying these equations constitutes a solution in semiclassical gravity: such a solution models a curved spacetime which incorporates quantum effects due to the propagation of the matter field, and it is expected to describe physical phenomena where the interplay between quantum matter and curvature becomes significant, like in the early Universe or in the vicinity of black holes. On the other hand, any semiclassical model represents an approximation of a more fundamental theory of quantum gravity, and thus it should be only valid in a regime where quantum gravity effects are negligible, and when fluctuations of the stress-energy tensor $\wick{T_{\mu\nu}}$ are small.

A conserved quantum stress-energy tensor which can enter \Eqs \eqref{eq:SCE} can be constructed in perturbative interacting quantum field theory in curved spacetimes according to the Hadamard point-splitting procedure \cite{Moretti03stress,Hollands05stress,Hack16cosm}. For the stress-energy tensor of a free Klein-Gordon field
\[
	T_{\mu\nu} = \nabla_\mu \phi \nabla_\nu \phi - \frac{1}{2} g_{\mu\nu} \at \nabla_\rho\phi \nabla^\rho \phi + m^2 \phi^2 \ct+ \xi \left(G_{\mu\nu} \phi^2 -\nabla_\mu\nabla_\nu\phi^2 + g_{\mu\nu} \nabla_\rho\nabla^\rho \phi^2 \right),
\]
the Hadamard renormalization procedure yields the following expectation value: 
\[
	\expvalom{T_{\mu\nu}}(x) = \lim_{x' \to x} {D_{\mu\nu} w(x',x)},
\]
where
\begin{align*}
	D_{\mu\nu} = &(1-2\xi) g^{\mu'}_\mu \nabla_{\mu'} \nabla_{\nu} - 2\xi \nabla_\mu\nabla_\nu + \xi G_{\mu\nu} + \\
	& g_{\mu\nu} \ag 2\xi \square_g  + \at 2\xi - \frac{1}{2} \ct g^{\rho'}_\rho \nabla_{\rho'}\nabla^{\rho} - \frac{1}{2} m^2 \cg + \frac{1}{3} g_{\mu\nu} P
\end{align*}
is the bi-differential operator which realizes the point-splitting of $T_{\mu\nu}$, and $w(x',x)$ is the smooth part of the Hadamard state $\omega$. In this expression, unprimed and primed indices denote covariant derivatives in the points $x$ and $x'$, respectively, while $g^{\nu'}_\nu$ is the bitensor of parallel transport. The last term proportional to $g_{\mu\nu}$, which vanishes in the classical case, must be added to obtain $\nabla^\mu \expvalom{T_{\mu\nu}}(x) = 0$. However, any construction of a covariantly conserved expectation value of $\wick{T_{\mu\nu}}$ leads to the violation of the classical conformal invariance of $T_{\mu\nu}$, and it gives rise to the so-called trace anomaly; namely, the trace of the quantum stress-energy tensor does not vanish for a massless, conformally coupled field, i.e., when $m =0$ and $\xi = 1/6$, different from its classical counterpart. This anomalous trace does not depend on the choice of the quantum state, but it is a local contribution depending on the geometry and on the linear equation of motion of the matter field. In the Hadamard point-splitting regularization scheme and for a free scalar field $\phi$, it is proportional to the Hadamard coefficient $v_1$ evaluated in the coinciding limit, and it reads in four dimensions as
\begin{equation}
\label{eq:trace}
	\expvalom{{T_{\rho}}^{\rho}} = \lambda \at C_{\alpha\beta\gamma\delta}C^{\alpha\beta\gamma\delta} + {R}_{\mu\nu}{R}^{\mu\nu} - \frac{1}{3}R^2 \ct, \qquad \lambda =\frac{1}{2880\pi^2},
\end{equation}
where $C_{\alpha\beta\gamma\delta}$ is the Weyl tensor, $R_{\mu\nu}$ is the Ricci tensor, and $R$ is the Ricci scalar. The $\square_g R$ usually appearing in $\expvalom{{T_{\rho}}^{\rho}}$ was removed by a judicious choice of the renormalization constants, according to the previous discussion about local and covariant Wick observables in curved spacetimes. In the case of $\expvalom{{T_{\rho}}^{\rho}}$, the renormalization scalar freedom is \cite{Moretti03stress,Hollands05stress}
\[
	Q = \tilde{\beta}_1 m^4 +  \tilde{\beta}_2 m^2 R + \tilde{\beta}_3 \square_g R,
\]
where $\tilde{\beta}_i \in \mathds{R}$ are renormalization constants, whose values can be changed after changing the length scale $\mu$ in the Hadamard parametrix \eqref{def:Hadamard}. Hence, in some special cases like the massless, conformally coupled field, the higher-order derivative term proportional to $\square_g R$ can be always cancelled by choosing properly $\tilde{\beta}_4$, thus meeting the requirement that curvature terms do not contain non-classical higher order derivatives of the metric \cite{Wald77back}.

\section{Evaporation of spherically symmetric black holes}

A model of evaporation of black holes which incorporates the back-reaction of the matter fields on the geometry can be fully developed in the framework of four-dimensional spherically symmetric spacetimes. In this class of geometries, invariants like the mass of the black hole and the area of the horizon can be provided thanks to the spherical symmetry \cite{Hayward98first}: the area of each two-sphere is $\cA = 4\pi r^2$, where $r$ is the radius, while the mass 
\[
	m = \frac{r}{2} \at 1 - \nabla^\rho r \nabla_\rho r \ct
\]  
represents the Misner-Sharp energy enclosed inside the sphere. The metric of every spherically symmetric spacetime can be always represented in double-null coordinates with respect to the null normal directions $\pa_V$ and $\pa_U$ as
\begin{equation}
\label{eq:double-null}
	\d s^2 = - 2A(V,U) \d V \d U + r^2(V,U) \d \Omega,
\end{equation}
where $\d \Omega = \d \theta^2 + \sin^2\theta \d \varphi^2$ denotes the line-element of the two-sphere of unit radius $\mathds{S}^2$. The orientation of the spacetime can be fixed in such a way that $A(V,U) > 0$ and at spatial infinity $\pa_V r >0$, $\pa_U r < 0$. To study evaporation, another convenient choice of parametrization is the Bardeen-Vaidya metric \cite{Bardeen81black}
\begin{equation}
\label{eq:Bardeen-Vaidya}
	\d s^2 = -\e^{2\Psi(v,r)} \at 1 - \frac{2m(v,r)}{r} \ct \d v^2 + 2\e^{\Psi(v,r)} \d v \d r + r^2 \d \Omega, 
\end{equation}
where $v$ denotes the advanced time coordinate and parametrizes any ingoing radial null geodesic. 

The notion of horizon in spherically symmetric black holes is encoded in the idea that no light rays can escape from this hypersurface, whereas they can fall inside. Namely, it is a  future outer local trapping horizon \cite{Hayward94dyn}. To formulate this idea on mathematical grounds, one introduces a pair of null radial geodesic congruences with respect to the outgoing and ingoing directions $\pa_V$ and $\pa_U$, respectively, and the relative expansion parameters
\[
	\theta_+ = \frac{2}{A r} \pa_V r, \qquad \theta_- = \frac{2}{Ar} \pa_U r.
\]
Each expansion parameter measures the rate of variation of the cross-sectional area of the geodesic congruence: $\theta_\pm = A^{-1} \pa_{V,U} \log(\cA)$, and it determines if the congruence is expanding or focusing along that null direction. Thus, the apparent horizon of a spherically symmetric black hole is defined as a three-manifold $\cH$ foliated by closed surfaces such that
\begin{equation}
\label{def:FOTH}
	\theta_+ \eqH 0, \qquad \theta_- \leqH 0, \qquad \pa_U \theta_+ \leqH 0,
\end{equation}
and it is located in $\cH : r - 2m = 0$  (the subscript $\cH$ labels the evaluation on the apparent horizon). The third condition captures the property that this trapping horizon is outer and it is referred to a black hole. In coordinates $(v,r)$, the hypersurface $\cH$ is a one-dimensional curve $r_\cH(v)$ in the plane $(v,r)$, and hence the mass of an evaporating black hole is defined by the relation
\begin{equation}
\label{eq:M}
	M(v) = m\at v,r_\cH(v) \ct = \frac{r_\cH(v)}{2}. 
\end{equation}

On spherically symmetric black holes, the mechanism of evaporation can be described in terms of the variation of the mass
\begin{equation}
\label{eq:DM}
\Delta M = \int_{\delta \cH} \d m
\end{equation}
along a small portion $\delta \cH$ of the apparent horizon (in the Bardeen-Vaidya metric \eqref{eq:Bardeen-Vaidya}, $\delta \cH$ corresponds to the curve enclosed between two points $P = (v_P,r_\cH(v_P))$ and $Q = (v_Q,r_\cH(v_Q))$ in the $(v,r)$ plane). If one considers a generic stress-energy tensor $T_{\mu\nu}$, then the evolution of $M(v)$ is governed by the $vv$-component of the (semiclassical) Einstein equations evaluated on $\cH$, which reads 
\begin{equation}
\label{eq:TVV-mass}
	\dot{M}(v) = \cA_\cH (v) {T_{vv}}(v,r_\cH(v)),
\end{equation}
where $\dot{M}(v) = \pa_v M(v)$ and $\cA_\cH = 4\pi r^2_\cH$ denote the rate of evaporation and the area of the horizon, respectively. Thus,
\begin{equation}
\label{eq:dM-Tvv}
	\Delta M = 4\pi \int_{v_P}^{v_Q} \frac{M}{\kappa} {T_{vv}}(r_\cH) \d v.
\end{equation}
As ${T_{vv}} \eqH {T_{VV}}$, both $\dot{M}(v)$ and $\Delta M$ depend on the ingoing flux of matter which is falling inside the black hole from the horizon. Whenever quantum matter is involved in the back-reaction process, it may happen that the quantum stress-energy tensor $\expvalom{T_{\mu\nu}}$ violates the classical null energy condition on the horizon, and hence the ingoing quantum component $\expvalom{T_{VV}}$ may be negative on $\delta \cH$. In the end, the presence of a negative ingoing flux on the horizon makes the black hole mass to evaporate with a negative rate $\dot{M}(v)< 0$.

\subsection{The static black hole}

The vacuum spherically symmetric spacetime is the Schwarzschild solution of the Einstein equations according to Birkhoff's theorem, and it describes a static spherical black hole of constant mass $M$ \cite{Wald84gr}. A double-null parametrization for arbitrary $U,V > 0$ is the maximally extension of the Schwarzschild eternal black hole in the so-called Kruskal-Szekeres coordinates
\[
	\d s^2 = - \frac{32 M^3}{r} \e^{-r/2M} \d V \d U + r^2(V,U) \d \Omega,
\]
where $r(V,U)$ is implicitly defined by 
\[
	UV = \at 1 - \frac{r}{2M} \ct \e^{r/2M}.
\]
On the other hand, the Bardeen-Vaidya metric corresponds to the advanced Eddington-Finkelstein coordinates, with $\Psi(v,r) = 0$ and $m(v,r) = M$.

\begin{figure}[ht] \centering{\includegraphics[scale=0.7]{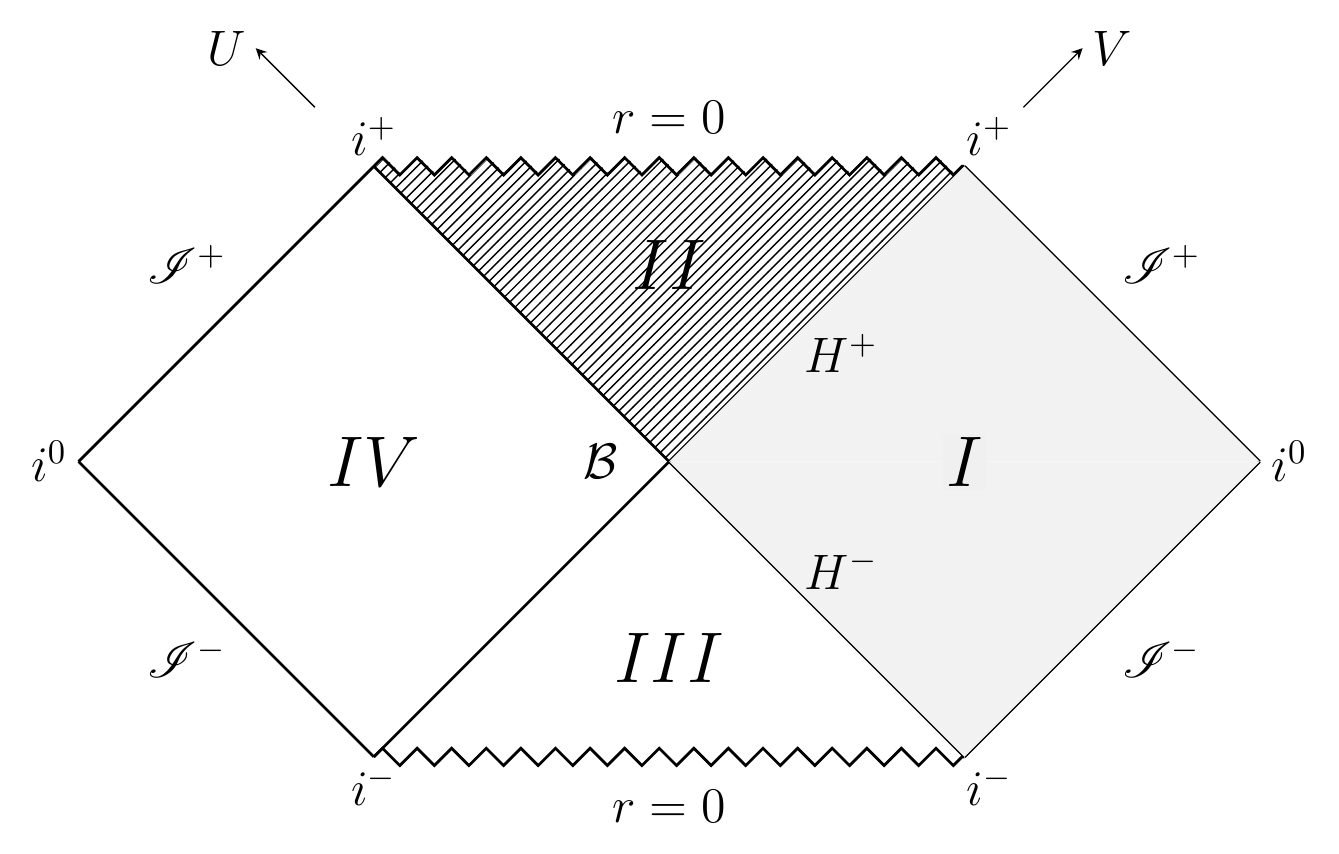}}
	\caption{Penrose diagram of the maximally extended Schwarzschild spacetime. Regions $I$ and $II$ are obtained in a static gravitational collapse, and they correspond to the outer region (grey region) and the inner region (dashed region) of the black hole, respectively. Region $III$ is the time-reversal counterpart of region $II$, and it describes a white hole; finally, region $IV$ is another asymptotically flat part of the spacetime, and it is connected to region $I$ by a wormhole. The bifurcated Killing horizon $H = H^+ \cup H^-$ associated to $\xi$ is composed by two null surfaces $H^{\pm}$ located in $U =0$ and $V = 0$, respectively, which intersect each other in the sphere of bifurcation $\mathcal{B}$; the event horizon corresponds to the future branch $H^+$ of $H$.}
	\label{aba:Kruskal}
\end{figure}

The static symmetry of the Schwarzschild spacetime reflects the existence of a Killing vector field $\xi = \frac{1}{4M} (V \pa_V - U \pa_U)$, which is unit timelike at infinity and it is hypersurface orthogonal. Due to the static symmetry, the apparent horizon coincides with the null event horizon of the black hole, which is located in the Schwarzschild radius $r = 2M$ (see Figure \ref{aba:Kruskal}\footnote{Penrose diagrams were made in TikZ and are based on the Latex code posted in StackExchange \cite{stack}.}). 

The surface gravity of the black hole is defined by the equation $\nabla^\mu \xi^2 = -2\kappa\xi^\mu$ and it constant along each null generator of $\xi$. Physically, the surface gravity represents the gravitational acceleration detected along the black hole horizon, and it is strictly positive on the event horizon.

According to Hawking's works, a static black hole can emits thermal radiation at infinity in form of a flux of quantum particles, if the back-reaction is not taken into account, and after assuming a vacuum state in the asymptotic past. The temperature of the radiation is the so-called Hawking temperature and it is proportional to the surface gravity of the black hole: in the units convention adopted in this paper, it reads
\[
	T_{H} = \frac{\kappa}{2\pi} = \frac{1}{8\pi k_B M},
\]
where $k_B$ is the Boltzmann constant. The correspondence between the temperature of the radiation and the surface gravity lets to interpret the first law of black hole mechanics
\[
	\d M = \frac{\kappa}{8\pi} \d\cA + \Omega \d J
\]
from a thermodynamic point of view, with $M$ and $\cA/4$ playing the role of energy and entropy of the black hole, respectively. More recently \cite{Fredenhagen90haw}, it was shown that in case of a static black hole formed in a gravitational collapse, the Hawking radiation with temperature $T_H$ always appears at future infinity whenever the state $\omega$ is a Hadamard state approaching the ground state near spatial infinity; for instance, in the case of the Unruh state \cite{Unruh76eva,Dappiaggi09unr}, which is regular on the horizon and respects the spherical symmetries of the spacetime.

Thus, the power of the radiation emitted from the black hole at future infinity is regarded as describing the evaporation of the black hole, because it can be instantaneously equated to the rate of loss of mass in the adiabatic approximation, namely neglecting how the spacetime is dynamically changing during the evaporation. Starting from this assumption, the negative ingoing flux on the horizon, that is, $\expvalom{T_{vv}} \eqH \expvalom{T_{VV}}$, can be related to the positive outgoing flux $\expvalom{T_{uu}}$ detected at infinity, where $u$ denotes the retarded Eddington-Finkelstein coordinate parametrizing outgoing null geodesics. In the Unruh state, it holds that 
\begin{equation}
\label{eq:flux-Schwarzschild}
	\frac{\dot{M}(v)}{4\pi} = \lim_{r \rightarrow r_{\cH}} \at \expvalom{T_{VV}} r^2 \ct = - \lim_{r \rightarrow \infty} \at \expvalom{T_{uu}} r^2 \ct = - \frac{L_\cH}{4\pi},
\end{equation}
where $L_\cH > 0$ denotes the luminosity of the black hole, namely the radial flux of particles measured by a static observer at large distance from the black hole \cite{Candelas80pol}. Indeed, \Eq \eqref{eq:flux-Schwarzschild} loses meaning outside the adiabatic approximation, since the rate $\dot{M}(v)$ always vanishes in case of constant mass. It turns out, in fact, that the Schwarzschild spacetime is never a solution of the semiclassical Einstein equations, and hence an eternal black hole cannot be in equilibrium with the back-reaction of any quantum matter field in the semiclassical approach.

\begin{figure}[ht] \centering{\includegraphics[scale=0.75]{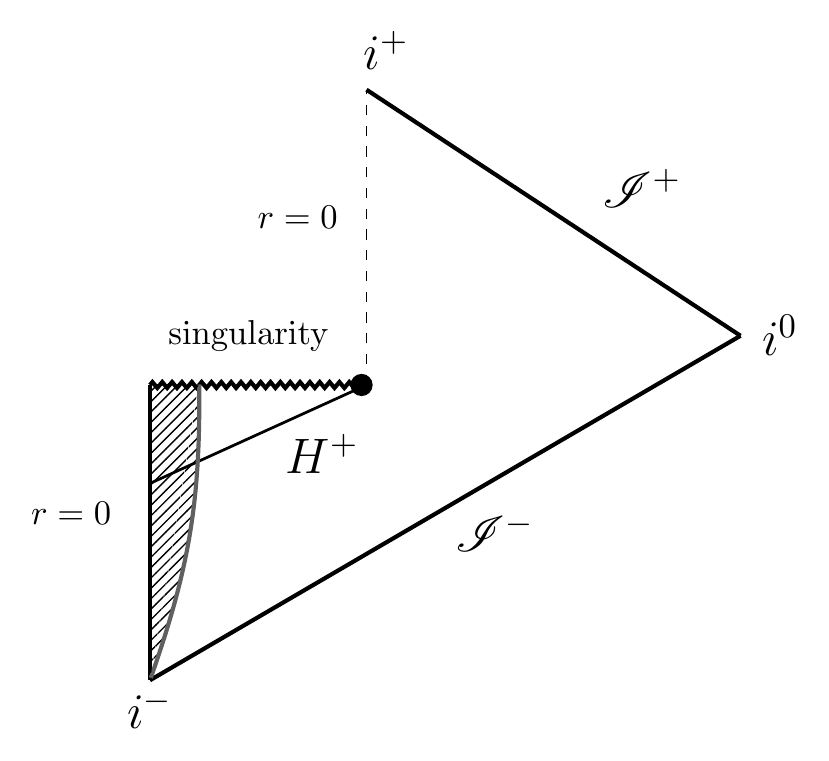}}
	\caption{Penrose diagram of a Schwarzschild gravitational collapse followed by evaporation and leaving an empty space at the end of the evolution. The black dot represents the end of the evaporation, and the dashed line denotes the new regular centre of the spacetime after the evaporation \cite{Hawking75creat}.}
	\label{aba:evap}
\end{figure} 

\subsection{Dynamical black holes}

The direct correspondence given in \Eq \eqref{eq:flux-Schwarzschild} between the ingoing negative flux and the positive outgoing flux holds in Schwarzschild spacetime thanks to the existence of a global Killing field which is timelike outside the event horizon, but it is missing in arbitrary dynamical backgrounds. In this case, the mechanism of evaporation and the following emission of power radiated should be dealt separately as causally related processes: on one hand, evaporation is a local effect induced by the negative ingoing flux and affecting the horizon; on the other hand, the expected Hawking radiation is associated to the positive outgoing flux emitted to infinity, and it depends globally on the entire history of spacetime.

Moreover, the adiabatic approximation is unable to describe evaporations when the spacetime is treated with its full dynamics, since it does not give informations about the evolution of the apparent horizon. In this respect, the Penrose diagram shown in Figure \ref{aba:evap} should be modified in case of dynamical backgrounds outside that approximation, because the dynamics of the horizon depends sharply on the negative flux $\expvalom{T_{VV}}$, and the causal structure of the spacetime may be significantly modified in the back-reaction process. Conversely, a semiclassical model of evaporation should necessarily rely on the semiclassical Einstein equations \eqref{eq:SCE} to study the variation of the mass $M(v)$ induced by the back-reaction of the quantum matter content, without further assumptions.
 
Contrary to Schwarzschild spacetimes modelling static black holes and event horizons, a general spherically symmetric spacetime describes a dynamical black hole and an apparent horizon, whose dynamics is dictated in \Eq \eqref{def:FOTH}. To define properties of spherically symmetric black holes like the surface gravity, there is a preferred choice of a vector field, which replaces the standard Killing vector field in stationary spacetimes. It is the Kodama vector\cite{Kodama80ko} $K = \text{curl}_n r$, where the curl is computed in the two-dimensional spacetime normal to the two-sphere. This vector is timelike in the region outside the horizon, it becomes lightlike on $\cH$, and eventually it is spacelike in the interior of the black hole. It also reduces to the Killing vector $\xi$ in stationary spacetimes. Furthermore, it is divergenceless: $\nabla_\mu K^\mu = 0$, and it can be used to define conserved currents $T_{\mu\nu} K^\nu$ given any symmetric tensor $T_{\mu\nu}$ which respects the spherical symmetry. Finally, from the definition of $K$ it holds that
\begin{equation}
	\mathscr{L}_K K_\mu = K^\nu \at \nabla_\nu K_\mu - \nabla_\mu K_\nu \ct \eqH \pm \kappa K_\mu,
\end{equation}
where $\mathscr{L}_K$ denotes the Lie derivative along $K$, and
\begin{equation}
\label{def:kodama-sg}
	\kappa \doteq \frac{1}{2} \square_n r,
\end{equation}
where the d'Alambertian operator is evaluated in the two-dimensional normal spacetime. Namely, this quantity corresponds to the definition of the surface gravity for a dynamical black hole, and it reduces to the standard one in the case of a Killing vector field. In particular, $\pa_U \theta_+ < 0$ implies that $\kappa > 0$ on $\cH$.

In analogy with the static case, Hayward showed \cite{Hayward98first} that the four laws of black hole mechanics can be generalized in presence of spherical symmetry to apparent horizons using the properties of the Kodama vector and the definition of the surface gravity given in  \Eq \eqref{def:kodama-sg}. In this class of spacetimes, the first law reads
\begin{equation}
\label{eq:first-law}
	m' = \frac{\kappa}{8\pi} \cA' + w V',
\end{equation}
where the prime denotes the derivative along the horizon, i.e., $f'= \ell \cdot \nabla f$, and $\ell$ is any tangent vector to $\cH$. Here, the second contribution in the right-hand side represents the variation of the work done by the matter on the horizon, with $V = \frac{4}{3} \pi r^3$ and $w = A^{-1} T_{UV}$. Furthermore, the same thermodynamic interpretation already seen for static black holes can be given in terms of the Kodama surface gravity: in this case, $\kappa/2\pi$ plays the role of dynamical local temperature of the apparent horizon with respect to the time parameter defined of the Kodama vector. This correspondence was shown in the dynamical case by studying the tunnelling probability of quantum matter across apparent horizons using the Hamilton-Jacobi method \cite{DiCriscienzo07tun,Vanzo11rev}, and by analyzing the scaling limit of the Hadamard two-point function on such horizons \cite{Kurpicz11temp}. 

\begin{figure}[ht] \centering{\includegraphics[scale=0.8]{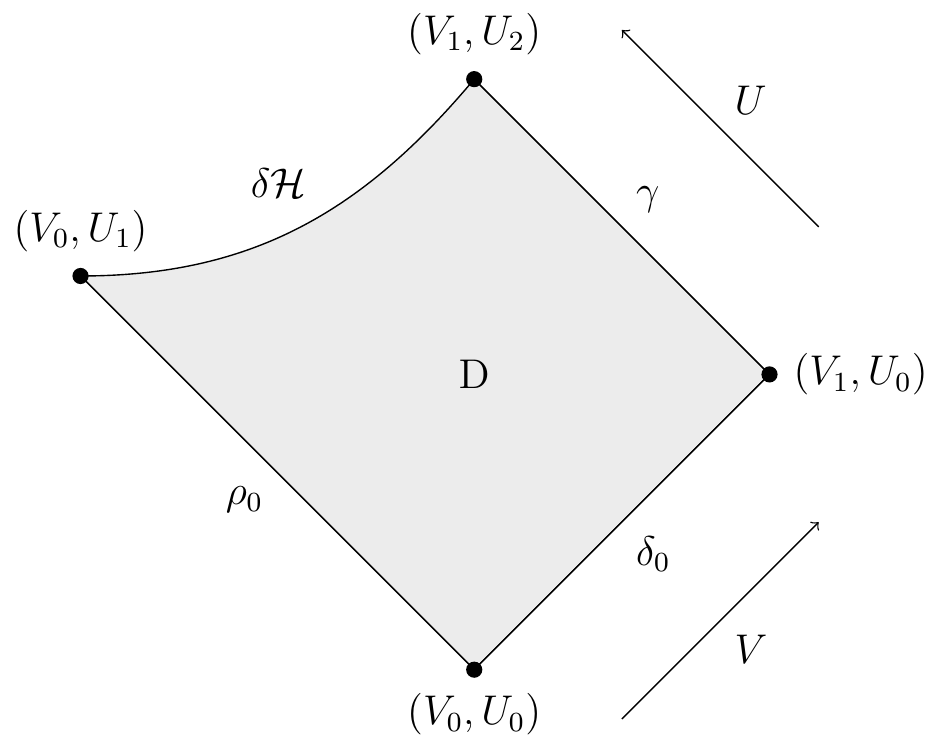}}
	\caption{Picture of the domain $D$. For spacelike $\delta \cH$, $\text{D} \times \mathds{S}^2 = J^{+}((V_0,U_0) \times \mathds{S}^2) \cap J^{-}(\delta \cH\times \mathds{S}^2)$, with $U_2 < U_1$. The points $P = (V_0,U_1)$ and $Q = (V_1,U_2)$ denote the extreme values of $\delta \cH$ in the $(V,U)$ plane. In describing evaporation, the initial conditions are posed on $S_0 = \{(V,U,\theta,\varphi)\in \mathcal{M} \mid V=V_0, \ U=U_0\}$ and on the curves $\rho_0$ and $\delta_0$ of the boundary $\partial \text{D}$\cite{Meda2021evap}.}
	\label{aba:D}
\end{figure}

According the previous section, \emph{cf.} \Eqs \eqref{eq:TVV-mass} and \eqref{eq:dM-Tvv}, evaporation of spherically symmetric black holes is fully determined by the ingoing flux located on the horizon. However, in the absence of further symmetries it is very challenging to evaluate the component $\expvalom{T_{VV}}$, and, more generally, the renormalized quantum stress-energy tensor $\expvalom{T_{\mu\nu}}$, in a state $\omega$ which was a vacuum state in the past. On the other hand, the variation of the mass of the black hole can be constrained by the matter-geometry content in the causal past and outside the black hole horizon \cite{Meda2021evap}. This is accomplished by applying the divergence theorem (Stokes' theorem) to the currents $J_r$ and $J_K$  obtained by contracting the stress-energy tensor with the gradient $\nabla r$ and with the Kodama vector $K$, respectively. The domain over which the divergence theorem is spherically symmetric and it has the form $\text{D} \times \mathds{S}^2$, where $\text{D}$ is a suitable region in the two-dimensional normal spacetime intersecting a portion of the apparent horizon. A representation of $\text{D}$ for a spacelike portion of the horizon is pictured in Figure \ref{aba:D}.

If one considers a stress-energy tensor $T_{\mu\nu}$ which respects the spherical symmetry and which satisfies vacuum-like initial condition at the boundary of $J^+(S_0)$, namely
\begin{equation}
\label{eq:initial-condition}
	T_{\mu\nu}(p) =0, \qquad p\in \partial J^+(S_0),
\end{equation}
then applying the divergence theorem to $J_r$ yields
\begin{equation}
\label{eq:dM-TUV-J}
	{\Delta M} = -\at \cS + \cW \ct,
\end{equation}
where
\begin{align}
	\cS &= 2\pi \int_{\text{D}} \nabla \cdot J_r \d V_\text{D} \label{eq:flux-S}, \\
	\cW &= 4\pi \int_{\gamma} \frac{T_{UV}r^2}{A} (-\pa_U r) \d U \label{eq:flux-W}.
\end{align}
Here, $\d V_\text{D} \doteq A r^2 \d V \wedge \d U$ is the volume form  on $\text{D}$ in the $(V,U)$ plane. \Eq \eqref{eq:flux-S} denotes the matter source inside the domain $\text{D}$ which influences the variation of the mass, while \Eq \eqref{eq:flux-W} is related to the work done by the matter when evaluated on $\cH$, in view of Hayward's first law \eqref{eq:first-law}. Moreover, applying the divergence theorem to $J_K$ in spacetimes where $\cW = 0$ yields
\begin{equation}
\label{eq:flux-W=0}
	\frac{\Delta M}{4\pi}\bigg\lvert_{{\footnotesize \cW = 0}} = - \int_{\gamma} \frac{T_{UU}r^2}{A} \pa_V r \d U.
\end{equation}
Namely, in the quantum case the variation of the mass is related to the outgoing flux $\expvalom{T_{UU}}$ across $\gamma$, which can interpreted as Hawking radiation sourced by $\cS$ and emitted from the evaporating black hole to infinity. In this view, the relation \eqref{eq:flux-W=0} generalizes \Eq \eqref{eq:flux-Schwarzschild} to more general non-static, spherically symmetric black holes outside the adiabatic approximation.

\subsection{The trace anomaly as source of evaporation}

It is known that the evaporation of a black hole, viewed as emission of Hawking radiation at infinity, can be ascribed to the presence of an anomalous trace in the stress-energy tensor of a quantum matter field in the vicinity of the horizon \cite{Davies76evap,Christensen77trace,Balbinot89back,Canfora03trace}. However, such analysis are often provided after making some approximations on the spacetime, like the adiabatic one. On the contrary, a full semiclassical approach to study the evaporation sourced by the trace anomaly should only rely on the semiclassical Einstein equations in four dimensions, without making further assumptions or referring to asymptotic effects which need global informations about the spacetime. 

Based on the results presented in the previous section, the variation of the black hole mass can be related to the source term $\cS$ given in \Eq \eqref{eq:flux-S}, and to the boundary term $\cW$ given in \Eq \eqref{eq:flux-W}. Thus, it was proved \cite{Meda2021evap} that the evaporation of a four-dimensional dynamical black hole can be induced by the trace anomaly \eqref{eq:trace} of a massless, conformally coupled scalar field affecting the horizon from the outside of the black hole, provided a weighted version of \Eq \eqref{eq:dM-TUV-J}. This result is obtained using only the semiclassical Einstein equations \eqref{eq:SCE}, once the initial condition \eqref{eq:initial-condition} is assumed on the state $\omega$ and a quantum averaged energy condition on $\expvalom{T_{UV}}$ is posed. Such a condition is of the form
\begin{equation}
\label{eq:AEC}
	\int_{U_0}^{U_\cH} \frac{\expvalom{T_{UV}} r^2}{A} f(V,U) A \d U \geq 0,
\end{equation}
where the integral is taken along any ingoing radial null curve connecting the initial point $(V,U_0)$ and $(V,U_\cH) \in \delta \cH$, and $f(V,U)$ is a certain strictly positive exponential function constructed from the geometry of the spacetime (its role is to tame the contributions inside $\expvalom{{T_{\rho}}^{\rho}}$ which do not contribute to a negative variation of the mass). The form of the function $f$ is similar to the ones entering usual quantum averaged weak energy conditions, which were shown to be valid both in flat and curved spacetimes for several values of the coupling-to-curvature parameter \cite{Kontou20qec}.

In principle, it is difficult to verify if such an energy condition can be fulfilled by the quantum collapsing matter during the evaporation process, because the lack of a quantum state in spherically symmetric spacetimes which was a vacuum in the past prevents to evaluate $\expvalom{T_{\mu\nu}}$ on and outside the horizon. However, one can expect that this condition is satisfied in an approximate way by a background geometry which is not too different from the one obtained in a classical model of gravitational collapse. Indeed, classical solutions are approximately valid in semiclassical gravity, because quantum corrections are usually small outside the horizon, and thus one can assume that solutions of the full semiclassical equations maintain their classical form outside the horizon. To provide a simple application, consider the Vaidya spacetime 
\[
	\d s^2 = -\at 1 - \frac{2M(v)}{r} \ct \d v^2 + 2 \d v \d r + r^2 \d \Omega, 
\]
which describes a null radiating spherically symmetric star with time-dependent mass $M(v)$. If one supposes that the quantum state fulfils $\expvalom{T_{UV}} = 0$, namely it makes the above metric a solution of \Eqs \eqref{eq:SCE}, then the negative ingoing flux on the horizon induced by the anomalous trace given in \Eq \eqref{eq:trace} can be evaluated explicitly, and it reads
\[
	\expvalom{T_{VV}} r^2 \eqH -\frac{3\lambda}{40M(v)^2}.
\]
Hence,
\[
	\dot{M}(v) = -\frac{3 \pi \lambda}{10M(v)^2}, \qquad \frac{\d S_\cH(v)}{\d v} = - \frac{12 \pi^2 \lambda}{5 M(v)},
\]
where $S_\cH = \cA_\cH / 4$ denotes the Wald-Kodama entropy.

\section{Conclusions}

The evaporation of four-dimensional spherically symmetric black holes in semiclassical gravity can be fully ascribed to the presence of a negative ingoing flux on the horizon, which determines the global behaviour of the decreasing mass of the black hole. Whereas the evaporation of any static black hole can be described directly in terms of Hawking radiation and black hole luminosity, for dynamical black holes the variation of the mass can be constrained by the quantum matter content outside and in the causal past of the black hole, without directly referring to any expected radiation detected at infinity. When conformally coupled quantum scalar fields are involved, the evaporation can be sourced by the trace anomaly of the quantum stress-energy tensor, assuming vacuum-like initial conditions in the past, and an auxiliary quantum energy condition outside the horizon. It is, however, very challenging to extend these results to more generic quantum matter contents, due to the difficulties in evaluating the renormalized stress-energy tensor in a quantum state. In order to overcome this issue, a semiclassical analysis which takes advantage of some quantum energy conditions similar to the one presented in \Eq \eqref{eq:AEC} might be viable in the future.

\section*{Acknowledgments}

The author would like to thank S. Roncallo for his careful reading of an earlier version of this proceeding.

\bibliographystyle{ws-procs961x669}
\bibliography{MGproceedingMeda}

\end{document}